# Bosonization and Lie Group Structure

**Yuan K. Ha**

Department of Physics, Temple University, Philadelphia, Pennsylvania 19122 U.S.A.

Email: yuanha@temple.edu

**Abstract**. We introduce a concise quantum operator formula for bosonization in which the Lie group structure appears in a natural way. The connection between fermions and bosons is found to be exactly the connection between Lie group elements and the group parameters. Bosonization is an extraordinary way of expressing the equation of motion of a complex fermion field in terms of a real scalar boson in two dimensions. All the properties of the fermion field theory are known to be preserved under this remarkable transformation with substantial simplification and elucidation of the original theory, much like Lie groups can be studied by their Lie algebras.

## 1. Introduction

Two-dimensional quantum field theories are important because the physics they describe is important. They contain exact symmetries and are often completely integrable. Some have exact solutions obtained without the need of perturbation theory and virtual particles. In contrast, no four-dimensional quantum field theories are known to be exactly solvable and therefore they can only be investigated as perturbative field theories.

Bosonization is a mathematical procedure in which a fermion field theory is expressed in terms of a boson theory with substantial simplification [1]. It is a unique phenomenon in two-dimensions. Bosonization is a search of the self-similarity of a system even though the equations appear to be different. It is therefore a novel symmetry operation in quantum field theory. In bosonization, any symmetry discovered which is not explicitly present in the fermion Lagrangian is a quantum symmetry. In particular, we find that the connection between fermions and bosons is exactly the connection between Lie group elements and the group parameters of the diagonal generators in the theory.

## 2. Quantum Fields

Fermions and bosons are fundamental fields in quantum field theory. Fermions are fields which obey anti-commutation rules

$$\{\psi_\rho(x), \psi_\sigma^\dagger(y)\} = \delta(x-y)\delta_{\rho\sigma}$$
$$\{\psi_\rho(x), \psi_\sigma(y)\} = 0. \tag{1}$$

where $\rho, \sigma = +, -$ are the two components of the fermion $\psi(x)$ in two dimensions. Bosons are fields which obey the commutation rule

$$[\phi(x), \phi(y)] = 0. \tag{2}$$

The Lagrangian for the free Dirac fermion with mass $m$ is

$$L = i\overline{\psi}(x)\gamma^{\mu}\partial_{\mu}\psi(x) - m\overline{\psi}(x)\psi(x), \tag{3}$$

with gamma matrices defined below:

$$\gamma^{0} = \begin{bmatrix} 0 & -i \\ i & 0 \end{bmatrix}, \quad \gamma^{1} = \begin{bmatrix} 0 & i \\ i & 0 \end{bmatrix}, \quad \gamma^{5} = \begin{bmatrix} 1 & 0 \\ 0 & -1 \end{bmatrix} = -i\gamma^{0}\gamma^{1}. \tag{4}$$

In the Weyl basis, the fermion components are given in the form

$$\psi = \begin{pmatrix} \psi_{+} \\ \psi_{-} \end{pmatrix} = \begin{pmatrix} \psi_{R} \\ \psi_{L} \end{pmatrix}, \tag{5}$$

in which the left- and right-handed components are respectively,

$$\psi_{L} = \frac{1 - \gamma^{5}}{2}\psi, \quad \psi_{R} = \frac{1 + \gamma^{5}}{2}\psi. \tag{6}$$

We also introduce the two-dimensional spacetime metric $g^{\mu\nu}$ with components $g^{00} = -g^{11} = 1$. The Lagrangian for the free boson with mass $\mu$ is

$$L = \frac{1}{2}\partial_{\mu}\phi(x)\partial^{\mu}\phi(x) - \mu^{2}\phi^{2}(x). \tag{7}$$

### 3. Fermion Current As Dynamical Variable

The key to bosonization lies in the recognition of the central role of the energy and momentum tensor. In the current algebra approach to quantum field theory, it is known that fermion current can be successfully used as dynamical variable. For the massless free fermion, we define the conventional fermion current operator as

$$j^{\mu}(x) = \overline{\psi}(x)\gamma^{\mu}\psi(x), \tag{8}$$

in terms of which the energy-momentum tensor takes the elegant form [2]

$$T^{\mu\nu}(x) = \frac{\pi}{2}\left(j^{\mu}(x)j^{\nu}(x) + j^{\nu}(x)j^{\mu}(x) - g^{\mu\nu}j_{\lambda}(x)j^{\lambda}(x)\right). \tag{9}$$

For the massless free boson, the canonical energy-momentum tensor is

$$T^{\mu\nu}(x) = \frac{1}{2}\partial^{\mu}\phi(x)\partial^{\nu}\phi(x) + \frac{1}{2}\partial^{\nu}\phi(x)\partial^{\mu}\phi(x) - g^{\mu\nu}\frac{1}{2}\partial_{\lambda}\phi(x)\partial^{\lambda}\phi(x). \tag{10}$$

It is obvious that the two energy-momentum tensors have a parallel structure. In addition, the commutator structure of the fermion currents at equal times, given by

$$[j^{\mu}(x), j^{\nu}(y)] = -\frac{i}{\pi}\varepsilon^{\mu\nu}\frac{\partial}{\partial x}\delta(x-y), \tag{11}$$

with $\varepsilon^{01} = -\varepsilon^{10} = 1$, is also very similar to the canonical commutation rule for the boson

$$[\phi(x), \pi(y)] = i\delta(x-y), \tag{12}$$

with $\pi(x) = \partial_{0}\phi(x)$ being the canonical conjugate of $\phi(x)$. The two energy-momentum tensors in equation (9) and equation (10), and the two commutators in equation (11) and equation (12) are identical, respectively, if the fermion current can be expressed as

$$j^{\mu}(x) = \frac{1}{\sqrt{\pi}}\varepsilon^{\mu\nu}\partial_{\nu}\phi(x). \tag{13}$$

This expression guarantees the fermion current conservation

$$\partial_{\mu}j^{\mu}(x) = 0 \tag{14}$$

due to the antisymmetry property of the tensor $\varepsilon^{\mu\nu}$. In this case, a single massless fermion has the same Hamiltonian as that for a single massless boson. We shall find that equation (13) is a crucial step in the construction of all bosonization.

### 4. Heisenberg Equation

The dynamics of the fermion field is governed by the Heisenberg equation of motion in quantum theory,

$$[P_{\mu}, \psi(x)] = -i\partial_{\mu}\psi(x), \tag{15}$$

in which the momentum operator $P_{\mu}$ is defined through the energy-momentum tensor from equation (9),

$$P^{\mu} = \int T^{0\mu}(x')dx',$$
$$T^{0\mu} = \frac{\pi}{2}(j^{0}j^{\mu} + j^{\mu}j^{0} - g^{0\mu}j_{\lambda}j^{\lambda}). \tag{16}$$

Applying the commutation relation between the current operator and the fermion at equal times,

$$[j^{\mu}(x), \psi(y)] = -(g^{0\mu} + \varepsilon^{0\mu}\gamma^{5})\psi(x)\delta(x-y), \tag{17}$$

we obtain the field equations of motion

$$\partial_0 \psi(x) = -i\pi[j^0(x) + \gamma^5 j^1(x)]\psi(x), \tag{18}$$

$$\partial_1 \psi(x) = +i\pi[j^1(x) + \gamma^5 j^0(x)]\psi(x). \tag{19}$$

Together they give the massless Dirac equation as required,

$$i\gamma^\mu \partial_\mu \psi(x) = 0. \tag{20}$$

We may formally integrate equation (19) by noting that the derivative of $\psi(x)$ is always proportionally to the field itself. This leads to an exponential expression for the fermion field of the form

$$\psi(x) = \exp\left[i\pi \int [j^1(x') + \gamma^5 j^0(x')]dx'\right]\psi_0, \tag{21}$$

where $\psi_0$ is a constant in the form of a spinor. According to equation (13), we may replace the current operators by the boson such that

$$j^0(x) = \frac{1}{\sqrt{\pi}} \partial_1 \phi(x), \qquad j^1(x) = -\frac{1}{\sqrt{\pi}} \partial_0 \phi(x). \tag{22}$$

The complete fermion expression in components then appears as

$$\psi(x) = \begin{bmatrix} \psi_+(x) \\ \psi_-(x) \end{bmatrix} = \begin{bmatrix} C_+ \exp\left[+i\sqrt{\pi}[\phi(x) - \int_{-\infty}^x \partial_0 \phi(x')dx']\right] \\ C_- \exp\left[-i\sqrt{\pi}[\phi(x) + \int_{-\infty}^x \partial_0 \phi(x')dx']\right] \end{bmatrix}, \tag{23}$$

with constants $C_+, C_-$ to be determined by the canonical anticommutation rule in equation (1). This is the explicit derivation of the formal result of bosonization in quantum field theory [3]. A complex fermion is thereby successfully constructed with a real scalar boson.

A remarkable simplification of equation (23) with profound significance can be achieved by introducing the chiral components of the boson

$$\phi_\pm(x) = \pm \frac{1}{\sqrt{2\pi}} \int_0^{\pm\infty} \frac{dk}{\sqrt{2k^0}} \left[c^{(-)}(k)e^{-ik\cdot x} + c^{(+)}(k)e^{ik\cdot x}\right] e^{-\alpha|k|/2}. \tag{24}$$

Here the scalar field $\phi(x)$ is decomposed into positive- and negative-energy parts: $\phi_+(x), \phi_-(x)$; $k^0$ is the energy frequency and $c^{(+)}(k), c^{(-)}(k)$ are the creation and annihilation operators of the boson with momentum $k$. The quantity $1/\alpha$ is an ultraviolet cutoff imposed to guarantee the boson to remain finite at high frequencies; $\alpha$ will be taken to zero at the end of any calculation involving the fermion $\psi(x)$. If the fermion components are to depend on the boson, each chiral component of the fermion should match perfectly to the chiral component of the boson with all energy frequencies in each component included. This is only possible if equation (23) takes the form

$$\begin{bmatrix} \psi_+(x) \\ \psi_-(x) \end{bmatrix} = \begin{bmatrix} C_+ \exp[+i\sqrt{4\pi}\phi_+(x)] \\ C_- \exp[-i\sqrt{4\pi}\phi_-(x)] \end{bmatrix}. \tag{25}$$

Accordingly, we find that

$$\phi_\pm(x) = \frac{1}{2}\left[\phi(x) \mp \int_{-\infty}^{x} \pi(x')dx'\right]. \tag{26}$$

The constants $C_+, C_-$ are evaluated and found to be equal to $1/\sqrt{2\pi\alpha}$. The fermion expression therefore takes the ultimate concise form

$$\psi(x) = \begin{bmatrix} \psi_+(x) \\ \psi_-(x) \end{bmatrix} = \frac{1}{\sqrt{2\pi\alpha}} \begin{bmatrix} \exp[+i\sqrt{4\pi}\phi_+(x)] \\ \exp[-i\sqrt{4\pi}\phi_-(x)] \end{bmatrix}. \tag{27}$$

This is the quintessential expression in the existing treatment of bosonization.

With the boson representation defined in equation (27), we find that bilinear terms in the fermion Lagrangian have their corresponding expressions in the following:

$$\begin{aligned}
i\overline{\psi}\gamma^\mu\partial_\mu\psi &\leftrightarrow \frac{1}{2}\partial_\mu\phi\partial^\mu\phi \\
\overline{\psi}\psi &\leftrightarrow -\frac{1}{\pi\alpha}\cos\sqrt{4\pi}\phi \\
i\overline{\psi}\gamma^5\psi &\leftrightarrow \frac{1}{\pi\alpha}\sin\sqrt{4\pi}\phi \\
\overline{\psi}\gamma^\mu\psi &\leftrightarrow \frac{1}{\sqrt{\pi}}\varepsilon^{\mu\nu}\partial_\nu\phi \\
\overline{\psi}\gamma^5\gamma^\mu\psi &\leftrightarrow \frac{1}{\sqrt{\pi}}\partial^\mu\phi.
\end{aligned} \tag{28}$$

These equivalences allow the bilinear terms appearing in the Lagrangian of a fermion theory to be transformed directly into the corresponding terms in the Lagrangian of a boson theory, while preserving all the key properties of the original fermion theory. Many symmetries in the fermion theory become immediately obvious in the bosonized form [4]. In addition, any hidden symmetry discovered which is not explicitly present in the fermion Lagrangian is a quantum symmetry. Quantum symmetries are symmetries acting on the field operators and they are typically discrete symmetries such as self-duality and self-triality [5].

## 5. Bosonization Theorem

Bosonization is an extraordinary way of expressing the equation of motion of a complex fermion field in term of a real scalar field with substantial simplification and elucidation of the original theory. Surprisingly, the bosonization formula given in equation (27) contains a natural Lie group structure. The exponential map from boson to fermion is strikingly similar to the exponential map from Lie algebra to

Lie group. Aside from the constant in equation (27), the fermion structure in terms of the boson is basically

$$\psi_+(x) = \exp+i\sqrt{4\pi}\phi_+(x)$$
$$\psi_-(x) = \exp-i\sqrt{4\pi}\phi_-(x). \quad (29)$$

For the $U(1)$ Lie group, the group elements of the unit circle are

$$z = \exp(+i\theta \cdot 1) = \cos\theta + i\sin\theta$$
$$z = \exp(-i\theta \cdot 1) = \cos\theta - i\sin\theta. \quad (30)$$

The quantity $\theta$ is a real number and it is the group parameter; $z$ is a complex number which is the group element. The identity $1$ is the group generator. For quantum fields, we note that the boson $\phi(x)$ is a real-valued function, whereas the fermion $\psi(x)$ is a complex-valued function. Thus in bosonization, if the boson is identified with the group parameter and the fermion with the group element, then the link between fermion and boson is exactly that given by the property of a Lie group. A subtlety, however, is to be noted. It is the chiral components which are independently bosonized. Each chiral component of the fermion is replaced by the corresponding chiral component of the boson and the complete fermion is reconstructed from these chiral components. This is necessary because the fermion is always a two-component spinor. We therefore arrive at a remarkable theorem in bosonization:

> **Theorem:** For any Lagrangian invariant under $U(1)$ symmetry transformation in two dimensions, if the chiral fermion is the group element, then the chiral boson is the group parameter and vice versa.

## 6. Interacting Fields

The power of bosonization is recognized by investigating a self-interacting theory, the massless Thirring model [6]. This is an exactly solvable quantum field theory with explicit solution to multi-point correlation functions. The Lagrangian is

$$L = i\overline{\psi}(x)\gamma^\mu \partial_\mu \psi(x) - \frac{g}{2} j^\mu(x) j_\mu(x). \quad (31)$$

For consistency with relativistic requirement, it is necessary that the fermion current in this theory be defined according to [7]

$$j^\mu(x) = \left(g^{0\mu} + \frac{\varepsilon^{0\mu}}{1+g/\pi}\right)\left(\overline{\psi}(x)\gamma^\mu \psi(x)\right), \quad (32)$$

with no summation over the index $\mu$. This is the current which is found to be conserved, i.e.

$$\partial_\mu j^\mu(x) = 0, \quad (33)$$

and therefore it can be expressed in terms of another boson $\tilde{\phi}(x)$ as

$$j^\mu(x) = \frac{1}{\sqrt{\pi}} \varepsilon^{\mu\nu} \partial_\nu \tilde{\phi}(x). \tag{34}$$

The bosonization formula in this interacting case is now given by

$$\psi(x) = \begin{bmatrix} \psi_+(x) \\ \psi_-(x) \end{bmatrix} = \frac{1}{\sqrt{2\pi\alpha}} \begin{bmatrix} \exp[+i\sqrt{4\pi}\tilde{\phi}_+(x)] \\ \exp[-i\sqrt{4\pi}\tilde{\phi}_-(x)] \end{bmatrix}, \tag{35}$$

with the chiral components of the boson being

$$\tilde{\phi}_\pm(x) = \frac{1}{2}\left( \tilde{\phi}(x) \mp \int_{-\infty}^{x} \tilde{\pi}(x')dx' \right)$$
$$\tilde{\phi}_+(x) + \tilde{\phi}_-(x) = \tilde{\phi}(x). \tag{36}$$

The new boson $\tilde{\phi}(x)$ and its canonical conjugate $\tilde{\pi}(x)$ reduce to free fields when the fermion coupling $g = 0$ as follow:

$$\tilde{\phi}(x) = \frac{\phi(x)}{\sqrt{1+g/\pi}}, \qquad \tilde{\pi}(x) = \sqrt{1+g/\pi} \cdot \partial_0 \phi(x). \tag{37}$$

The bosonized Lagrangian from equation (31) is therefore

$$\begin{aligned} L &= \frac{1}{2}\partial_\mu \tilde{\phi}(x)\partial^\mu \tilde{\phi}(x) + \frac{g}{2}\left( \frac{1}{\pi}\partial_\alpha \tilde{\phi}(x)\partial^\alpha \tilde{\phi}(x) \right) \\ &= \left(1 + \frac{g}{\pi}\right) \times \frac{1}{2} \frac{\partial_\mu \phi(x)}{\sqrt{1+g/\pi}} \frac{\partial^\mu \phi(x)}{\sqrt{1+g/\pi}} \\ &= \frac{1}{2}\partial_\mu \phi(x)\partial^\mu \phi(x). \end{aligned} \tag{38}$$

The massless Thirring model is immediately found to be equivalent to a theory of free massless boson as confirmed by examining the correlation functions and other key field theory properties. The chiral fermions and the chiral bosons in the interacting case also conform to a Lie group structure.

For the massive Thirring model, the result of bosonization gives the quantum sine-Gordon theory [8]. The result is consistent with the known solution of the mass spectrum and the scattering matrix in the massive Thirring model. The fundamental fermion of the Thirring model corresponds to the soliton of the sine-Gordon theory. The Lagrangians are, respectively,

$$\begin{aligned} L &= i\bar{\psi}(x)\gamma^\mu \partial_\mu \psi(x) - \frac{g}{2} j^\mu(x) j_\mu(x) - m\bar{\psi}(x)\psi(x), \\ L &= \frac{1}{2}\partial_\mu \phi(x)\partial^\mu \phi(x) + \frac{m}{\pi\alpha} \cos\sqrt{\frac{4\pi}{1+g/\pi}}\phi(x). \end{aligned} \tag{39}$$

## 7. U(2) Thirring Model

We consider an interacting theory involving two massless fermion species $\psi^1(x), \psi^2(x)$, each of which is a two-component spinor. They form an isovector under global $U(2)$ symmetry transformation. The model is important in the study of many two-dimensional systems and the Kondo problem [9]. There are four chiral components interacting according to the Thirring Lagrangian

$$L = i\bar{\psi}^1(x)\gamma^\mu \partial_\mu \psi^1(x) + i\bar{\psi}^2(x)\gamma^\mu \partial_\mu \psi^2(x) - gJ^{A,\mu}(x)J^A_\mu(x). \tag{40}$$

The isospin currents are defined as

$$J^{A,\mu}(x) = \left(g^{0\mu} + \frac{\varepsilon^{0\mu}}{1+g/\pi}\right)\bar{\psi}^a(x)\gamma^\mu \frac{\lambda^A}{2}\psi^a(x), \tag{41}$$

with the indices $a = 1,2$ and $A = 0,1,2,3$. The generators for the $U(2)$ Lie group are the following:

$$\lambda^0 = \begin{bmatrix} 1 & 0 \\ 0 & 1 \end{bmatrix}, \quad \lambda^1 = \begin{bmatrix} 0 & 1 \\ 1 & 0 \end{bmatrix}, \quad \lambda^2 = \begin{bmatrix} 0 & -i \\ i & 0 \end{bmatrix}, \quad \lambda^3 = \begin{bmatrix} 1 & 0 \\ 0 & -1 \end{bmatrix}. \tag{42}$$

They satisfy the $SU(2)$ commutation relations

$$\left[\frac{\lambda^i}{2}, \frac{\lambda^j}{2}\right] = i\varepsilon_{ijk}\frac{\lambda^k}{2}, \qquad i,j,k = 1,2,3. \tag{43}$$

The structure constants are the totally antisymmetric tensor $\varepsilon_{ijk}$ with $\varepsilon_{123} = 1$. The dynamics of the $U(2)$ Thirring model shows that only the currents corresponding to the diagonal generators are conserved and they can be expressed in terms of two real scalar fields $\tilde{\phi}^1(x), \tilde{\phi}^2(x)$,

$$\partial_\mu J^{A=0,\mu}(x) = 0 \longrightarrow J^{A=0,\mu} = \frac{1}{\sqrt{\pi}}\varepsilon^{\mu\nu}\partial_\nu \tilde{\phi}^1(x)$$

$$\partial_\mu J^{A=3,\mu}(x) = 0 \longrightarrow J^{A=3,\mu} = \frac{1}{\sqrt{\pi}}\varepsilon^{\mu\nu}\partial_\nu \tilde{\phi}^2(x). \tag{44}$$

Both scalar fields reduce to free fields in the non-interacting case analogous to equation (37).
 The bosonization formula in the $U(2)$ Thirring model therefore becomes

$$\psi^a(x) = \begin{bmatrix} \psi^a_+(x) \\ \psi^a_-(x) \end{bmatrix} = \frac{1}{\sqrt{2\pi\alpha}}\begin{bmatrix} \exp[+i\sqrt{4\pi}\tilde{\phi}^a_+(x)] \\ \exp[-i\sqrt{4\pi}\tilde{\phi}^a_-(x)] \end{bmatrix}, \tag{45}$$

with each fermion $\psi^a(x)$ corresponding to its own boson $\phi^a(x)$, $a = 1,2$. Again, we find that in this interacting case there is an underlying Lie group structure between the chiral components of the fermions and the chiral components of the bosons as a consequence of the equation of motion.

The energy-momentum tensor for the $U(2)$ Thirring model is appropriately defined as

$$T^{\mu\nu}(x) = \left(1 + \frac{g}{\pi}\right)\frac{\pi}{2}\sum_{A=0}^{3} J^{A,\mu}(x)J^{A,\nu}(x) + J^{A,\nu}(x)J^{A,\mu}(x) - g^{\mu\nu}J^{A}{}_{\lambda}(x)J^{A,\lambda}(x), \qquad (46)$$

from which the Hamiltonian $T^{00}(x)$ is calculated. In terms of ordinary scalar field notation, we find

$$T^{00}(x) = \frac{1}{2}(\partial_0\phi^1)^2 + \frac{1}{2}(\partial_1\phi^1)^2 + \frac{1}{2}(\partial_0\phi^2)^2 + \frac{1}{2}(\partial_1\phi^2)^2 - \frac{g}{4\pi^2\alpha^2}\cos\sqrt{\frac{4\pi}{1+g/\pi}}(\phi^1 - \phi^2). \qquad (47)$$

The Lagrangian which yields the above Hamiltonian is therefore

$$L = \frac{1}{2}\partial_\mu\phi^1\partial^\mu\phi^1 + \frac{1}{2}\partial_\mu\phi^2\partial^\mu\phi^2 + \frac{g}{4\pi^2\alpha^2}\cos\sqrt{\frac{4\pi}{1+g/\pi}}(\phi^1 - \phi^2). \qquad (48)$$

By defining the following linear combinations:

$$A = (\phi^1 + \phi^2)/\sqrt{2}, \qquad B = (\phi^1 - \phi^2)/\sqrt{2}, \qquad (49)$$

we may decouple a free massless boson $A(x)$ and obtain a sine-Gordon theory with boson $B(x)$ through the Lagrangian

$$L = \frac{1}{2}\partial_\mu A\partial^\mu A + \left(\frac{1}{2}\partial_\mu B\partial^\mu B + \frac{g}{4\pi^2\alpha^2}\cos\sqrt{\frac{8\pi}{1+g/\pi}}B\right). \qquad (50)$$

This is the decomposition of the Lie algebra $u(2) \approx u(1) \otimes su(2)$ in terms of the bosons and the realization of the isomorphism of the group $U(2) \approx U(1) \otimes SU(2)$ in terms of the fermions. The sine-Gordon Lagrangian in equation (50) is further equivalent to another massive Thirring model represented by a fermion $\chi(x)$ of mass $m'$ with Lagrangian

$$L = i\bar{\chi}(x)\gamma^\mu\partial_\mu\chi(x) - \frac{g'}{2}k^\mu(x)k_\mu(x) - m'\bar{\chi}(x)\chi(x). \qquad (51)$$

The fermion current here is given by

$$k^\mu(x) = \left(g^{\prime 0\mu} + \frac{\varepsilon^{0\mu}}{1+g'/\pi}\right)(\bar{\chi}(x)\gamma^\mu\chi(x)). \qquad (52)$$

The conservation of this fermion current then allows it to be expressed in terms of another boson $\tilde{B}(x)$, i.e.

$$\partial_\mu k^\mu(x) = 0 \longrightarrow k^\mu(x) = \frac{1}{\sqrt{\pi}}\varepsilon^{\mu\nu}\partial_\nu\tilde{B}(x). \qquad (53)$$

The bosonization formula involved is,

$$\chi(x) = \begin{bmatrix} \chi_+(x) \\ \chi_-(x) \end{bmatrix} = \frac{1}{\sqrt{2\pi\alpha}} \begin{bmatrix} \exp[+i\sqrt{4\pi}\tilde{B}_+(x)] \\ \exp[-i\sqrt{4\pi}\tilde{B}_-(x)] \end{bmatrix}, \qquad (54)$$

where the fields $\tilde{B}(x)$ and $B(x)$ are related by

$$\tilde{B}(x) = \frac{B(x)}{\sqrt{1+g'/\pi}}, \qquad (55)$$

with the identification of the parameters $m' = g/4\pi\alpha$ and $g' = (g-\pi)/2$.

## 8. Dynamical Mass Generation

The $U(2)$ Thirring model with two massless fermions is thus seen to be equivalent to the single massive fermion $U(1)$ Thirring model via bosonization. The coupling constant $g$ in the $U(2)$ fermion theory becomes the mass parameter $m'$ in the $U(1)$ Thirring model. This is dynamical mass generation. The single fermion $\chi(x)$ in the massive Thirring model is a non-linear composite of the original Fermi fields $\psi^1(x), \psi^2(x)$ in the $U(2)$ massless case. It is feasible to generalize to the $U(N)$ Thirring model with $N$ massless fermions [10]. The fermion currents corresponding to the diagonal generators of the $U(N)$ group will be conserved, resulting in $N$ scalar fields to form a $N$ coupled sine-Gordon Lagrangian. Bosonization maintains all the symmetries of the original fermion theory and, in particular, reveals a natural connection between fermion and boson in terms of Lie group structure. It becomes a novel and important symmetry operation in quantum field theory, very much like Lie groups can be studied effectively by their Lie algebras.